\documentclass[a4paper,11pt]{article}

\usepackage{pos}

\title{
  Worldvolume tempered Lefschetz thimble method 
  and its error estimation%
  \footnote{Report No.: KUNS-2905}
}
\ShortTitle{Worldvolume tempered Lefschetz thimble method 
and its error estimation}

\author[a]{Masafumi Fukuma}
\author*[b]{Nobuyuki Matsumoto}
\author[a]{Yusuke Namekawa}

\affiliation[a]{
  Department of Physics, Kyoto University\\
  Kyoto 606-8502, Japan
}
\affiliation[b]{
  RIKEN/BNL Research center, Brookhaven National Laboratory,\\
  Upton, NY 11973, USA
}

\emailAdd{fukuma@gauge.scphys.kyoto-u.ac.jp}
\emailAdd{nobuyuki.matsumoto@riken.jp}
\emailAdd{namekawa@gauge.scphys.kyoto-u.ac.jp}

\abstract{
  The worldvolume tempered Lefschetz thimble method (WV-TLTM) 
  is an algorithm towards solving the sign problem, 
  where hybrid Monte Carlo updates are performed 
  on a continuous accumulation of flowed surfaces 
  foliated by the anti-holomorphic gradient flow 
  (the worldvolume of integration surface). 
  Sharing the advantage 
  with the original tempered Lefschetz thimble method (TLTM) 
  that the sign problem is resolved 
  without introducing the ergodicity problem, 
  the new algorithm is expected to significantly reduce the computational cost, 
  because it eliminates the need to compute the Jacobian of the flow 
  in generating a configuration. 
  We demonstrate the effectiveness of the WV-TLTM 
  with its successful application to the Stephanov model 
  (a chiral random matrix model), 
  for which the complex Langevin method is known 
  to suffer from a serious wrong convergence problem. 
  We also discuss the statistical analysis method for the WV-TLTM. 
}

\FullConference{%
 The 38th International Symposium on Lattice Field Theory, LATTICE2021
  26th-30th July, 2021
  Zoom/Gather@Massachusetts Institute of Technology
}

\newcommand{\sigt}{{\Sigma_t}}
\newcommand{\calr}{{\mathcal{R}}}
\newcommand{\calo}{{\mathcal{O}}}
\newcommand{\calj}{{\mathcal{J}}}
\newcommand{\rn}{{\mathbb{R}^N}}
\newcommand{\cn}{{\mathbb{C}^N}}
\newcommand{\bbR}{{\mathbb{R}}}
\newcommand{\bbC}{{\mathbb{C}}}
\newcommand{\re}{{\rm Re}\,}
\newcommand{\im}{{\rm Im}\,}

\begin{document}
\maketitle

\section{Introduction}
\label{sec:introduction}

The sign problem is a notorious problem 
in applying Monte Carlo methods to systems with complex actions, 
which include the finite density QCD as a typical example 
\cite{Guenther:2021}. 
The tempered Lefschetz thimble method (TLTM) \cite{Fukuma:2017fjq} 
was proposed as a versatile solution to this problem. 
This algorithm resolves the ergodicity problem 
inherent in Lefschetz thimble methods 
by tempering the system with the antiholomorphic gradient flow, 
where a discrete set of replicas of configuration space is introduced. 
In spite of its successful applications to various models 
\cite{Fukuma:2017fjq,Fukuma:2019wbv,Fukuma:2019uot}, 
the original TLTM has a drawback of high computational costs 
in generating a configuration, 
coming from the calculation of the Jacobian of the flow 
and from the introduction of a large number of replicas 
to ensure a sufficiently large acceptance rate 
in exchanging configurations between adjacent replicas. 

The {\it worldvolume tempered Lefschetz thimble method} (WV-TLTM) 
\cite{Fukuma:2020fez} 
was introduced to overcome this issue. 
This algorithm belongs to a class of 
\emph{worldvolume Hybrid Monte Carlo} (WV-HMC), 
where HMC updates are performed on a continuous set of replicas 
(the \emph{worldvolume} of integration surface).
The WV-TLTM is expected to significantly  reduce the computational costs, 
because one no longer needs to compute the Jacobian of the flow 
in generating a configuration. 
In this talk, 
we review the WV-TLTM 
and show its successful application \cite{Fukuma:2020fez} 
to the Stephanov model \cite{Stephanov:1996ki}. 
We also discuss the statistical analysis method 
for this algorithm \cite{Fukuma:2021aoo}.
The basics and other applications of the (WV-)TLTM  
are discussed in the contribution \cite{Fukuma:2021}.

\section{Worldvolume tempered Lefschetz thimble method}
\label{sec:worldv-temp-lefsch}

In the Lefschetz thimble method 
\cite{Witten:2010cx,Cristoforetti:2012su,Fujii:2013sra,Alexandru:2015sua}, 
we complexify the integration variable from $x\in\bbR^N$ 
to $z=x+i y\in\bbC^N$. 
We assume that 
the integrands in path integrals are entire functions over $\cn$. 
Then, Cauchy's theorem ensures that 
the integration surface can be continuously deformed, 
$\Sigma_0=\rn\to\Sigma_t$ ($t$\,: deformation parameter), 
without changing the values of the integrals: 
\begin{align}
  \langle \calo \rangle
  \equiv
    \frac{ \int_{\rn}dx\,e^{-S(x)} \calo(x) }
  { \int_{\rn}dx\,e^{-S(x)} }
  =
  \frac{ \int_{\sigt} dz_t\, e^{-S(z_t)} \mathcal{O}(z_t) }
  { \int_{\sigt} dz_t\, e^{-S(z_t)} }. 
\label{eq:1}
\end{align}
We deform the integration surface
with the following anti-holomorphic gradient flow:
\begin{align}
  \frac{d}{dt} z_t = [\partial S(z_t)]^*, \quad
  z_{t=0} = x.  
\label{eq:3}
\end{align}
The crucial property of the flow is 
that the flowed surface approaches a union of Lefschetz thimbles:
\begin{align}
  \sigt \xrightarrow{t\to\infty} \bigcup_\sigma \calj_\sigma, 
\end{align}
where the Lefschetz thimble $\calj_\sigma$ 
associated with a critical point $z_\sigma$  
[at which $\partial S(z_\sigma)=0$] 
consists of orbits flowing out of $z_\sigma$. 
Since the imaginary part of the action is constant 
on each $\calj_\sigma$ 
[$\im S(z) = \im S(z_\sigma)$ $(z \in \calj_\sigma)$], 
the oscillatory behavior of the integrals will become mild at large $t$. 
However, multiple $\calj_\sigma$'s are often relevant 
to the estimation of the path integral \cite{Fujii:2015bua}, 
and since $\calj_\sigma$'s are separated by regions where 
$\re S(z)=\infty$, 
the distribution on $\Sigma_t$ gets multimodal 
as we increase the flow time $t$. 

The tempered Lefschetz thimble method (TLTM) \cite{Fukuma:2017fjq} 
resolves this issue 
by tempering the system with the flow time 
(where the parallel tempering 
\cite{Swendsen1986,Geyer1991,Hukushima1996} is employed). 
However, this algorithm becomes expensive at large degrees of freedom 
because of the increasing number of replicas 
so as to ensure a sufficiently large acceptance rate 
in the exchange process, 
as well as of the calculation of the determinant 
of the Jacobian of the flow, 
$J_t(x) \equiv \partial z_t(x)/\partial x$.%
\footnote{
  $|\det J_t(x)|$ needs to be computed in the exchange process  
  to take into account the difference of the volume elements 
  between adjacent replicas
  (see, e.g., \cite{Fukuma:2019uot} for a detailed discussion). 
} 

\begin{figure}[t]
  \centering
  \includegraphics[width=80mm]{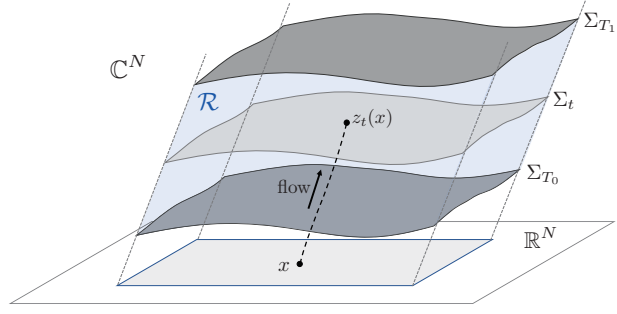}
  \caption{
    The worldvolume $\calr$ (colored region) consisting of $\sigt$ 
    foliated by the flow \eqref{eq:3} . 
  }
  \label{fig:wv}
\end{figure}

The worldvolume tempered Lefschetz thimble method (WV-TLTM) \cite{Fukuma:2020fez} 
resolves this drawback of the original TLTM. 
The key observation in the new algorithm is that 
Eq.~\eqref{eq:1} can be expressed as an integral over a region 
\begin{align}
  \calr 
  \equiv \bigl\{z_t(x)\in\bbC^N \,\bigl|\, \, x\in\rn,\, t\in[T_0, T_1]\bigr\} 
  = \bigcup_{t=T_0}^{T_1} \sigt, 
\end{align}
by averaging the numerator and the denominator in Eq.~\eqref{eq:1} 
over $t$ with an arbitrary weight $e^{-W(t)}$:%
\footnote{ 
  Although $e^{-W(t)}$ can take an arbitrary form in principle, 
  practically it is chosen 
  such that it leads to a uniform distribution in $t$.
} 
\begin{align}
  \langle \calo \rangle
  =
  \frac{ \int_{\sigt} dz_t\, e^{-S(z_t)} \mathcal{O}(z_t) }
  { \int_{\sigt} dz_t\, e^{-S(z_t)} } 
  =
  \frac{ \int_{T_0}^{T_1} dt\, e^{-W(t)}\int_{\sigt} dz_t\, 
  e^{-S(z_t)} \mathcal{O}(z_t) }
  {  \int_{T_0}^{T_1} dt\, e^{-W(t)}\int_{\sigt} dz_t\, e^{-S(z_t)} }. 
\label{eq:2}
\end{align}
We call the region $\calr$ the \emph{worldvolume}, 
because this can be regarded as an orbit of integration surface  
(see Fig.~\ref{fig:wv}). 
To perform Monte Carlo calculations, 
we further rewrite Eq.~\eqref{eq:2} to the form
\begin{align}
  \langle \calo \rangle
  =
  \frac{ \int_{\calr} Dz\, e^{-[W(t(z))+\re S(z)]}
  \bigl( dt\, dz_t / Dz \bigr)\,
  e^{-i\,\im S(z)}\, \mathcal{O}(z) }
  {  \int_{\calr} Dz\, e^{-[W(t(z))+\re S(z)]}
  \bigl( dt\, dz_t / Dz \bigr)\,
  e^{-i\,\im S(z)} }, 
\label{eq:mathcalO}
\end{align}
where $Dz$ is the invariant volume element on $\calr$ 
and $t(z)$ is the flow time at $z\in \calr$. 
If we use $(t,x)$ as coordinates on $\calr$, 
then $dz_t$ and $Dz$ can be written as 
\begin{align}
  dz_t &= \det J_t(x)\,dx,
\\
  Dz &= \alpha(z_t(x))\,|\det J_t(x)| \,dx\, dt, 
\end{align}
where $\alpha(z_t)$ is the magnitude of the normal component 
of $\dot{z}_t(x)=[\partial S(z_t(x))]^\ast$ to $\calr$. 
Then, by introducing the potential $V(z)$ and the reweighting factor $A(z)$ as 
\begin{align}
  V(z) &\equiv \re S(z) + W(t(z)), 
\label{eq:V}
\\
  A(z) &\equiv \frac{dt dz_t}{Dz}\, e^{-i\,\im S(z)}
  = \alpha^{-1}(z)\, \frac{\det\,J_t(x)}{|\det\,J_t(x)|}\, 
  e^{-i \, \im S(z)}, 
\label{eq:5}
\end{align}
Eq.~\eqref{eq:mathcalO} can be written as a ratio of reweighted averages, 
\begin{align}
  \langle \calo \rangle =
  \frac{\langle A(z)\,\calo(z) \rangle_\calr}
  {\langle A(z)\rangle_\calr}
  \quad
  \biggl(\langle f(z) \rangle_\calr \equiv 
  \frac{ \int_{\calr} Dz\, e^{-V(z)} f(z) }
  { \int_{\calr} Dz\, e^{-V(z)}} \biggr). 
\end{align}
The reweighted averages $\langle \,\ast\, \rangle_\calr$ on the subspace $\calr$ 
are estimated by the Hybrid Monte Carlo 
with the RATTLE discretization \cite{Andersen:1983,Leimkuhler:1994} 
(for the use of the RATTLE in Lefschetz thimble methods, 
see \cite{Fujii:2013sra,Alexandru:2019} and \cite{Fukuma:2019uot}).

The computational cost of the WV-TLTM is reduced 
from that of the original TLTM for two reasons. 
One is that 
the continuous migration of $t$ eliminates 
the need to worry about the acceptance rate in the exchange process. 
The other is that, 
as the potential $V(z)$ in the molecular dynamics 
does not contain the Jacobian matrix $J_t(x)$ [see Eq.~\eqref{eq:V}], 
$J_t(x)$ needs not be calculated in generating a configuration 
if we use an iterative method in linear inversion 
as in Ref.~\cite{Alexandru:2017lqr} 
(see Ref.~\cite{Fukuma:2020fez} for details).%
\footnote{ 
  The phase of the Jacobian [Eq.~\eqref{eq:5}] 
  still needs to be computed upon measurement. 
} 

\section{Application to the Stephanov model}
\label{sec:appl-steph-model}

To confirm the effectiveness of the WV-TLTM, 
we apply this algorithm to the Stephanov model \cite{Stephanov:1996ki}. 
This model has been regarded as an important benchmark 
for algorithms towards solving the sign problem, 
because it reproduces the qualitative behavior of 
finite density QCD in the limit of large matrix size 
\cite{Shuryak:1992pi,Stephanov:1996ki,Halasz:1998qr} 
and also it gives rise to a serious wrong convergence problem 
\cite{Bloch:2017sex}
with the complex Langevin method \cite{Parisi:1983cs,Klauder:1983sp}. 

The partition function of the model is given by 
\begin{align}
  Z_{\rm Steph} \equiv \int d^2 W e^{-n \, {\rm tr}\, W^{\dagger}W}
  {\rm det}
  \begin{bmatrix}
    m & i W + \mu  \\
    i W^\dagger + \mu  & m \\
  \end{bmatrix}.
\end{align}
Here, $W$ is an $n \times n$ complex matrix 
representing the gauge-field degrees of freedom 
with a flat measure $d^2W \equiv \prod_{i,j} (d\re W_{ij}\, d\im W_{ij})$, 
and $m$ and $\mu$ correspond 
to the quark mass and the baryon number chemical potential, respectively. 
We have set the temperature to zero, 
for which the sign problem is most severe.
We estimate the chiral condensate $\langle \overline{\psi} \psi\rangle$ 
and the baryon number density $\langle \psi^\dag \psi\rangle$
defined by
\begin{align}
  \langle \overline{\psi} \psi \rangle
  \equiv \frac{1}{2n}\,
  \frac{\partial}{\partial m} \log Z_{\rm Steph},\quad
  \langle \psi^\dagger \psi \rangle
  \equiv \frac{1}{2n}\,
    \frac{\partial}{\partial \mu} \log Z_{\rm Steph}.
\label{eq:chiral_condensate_number_density}
\end{align}

\begin{figure}[t]
  \centering
  \includegraphics[width=65mm]{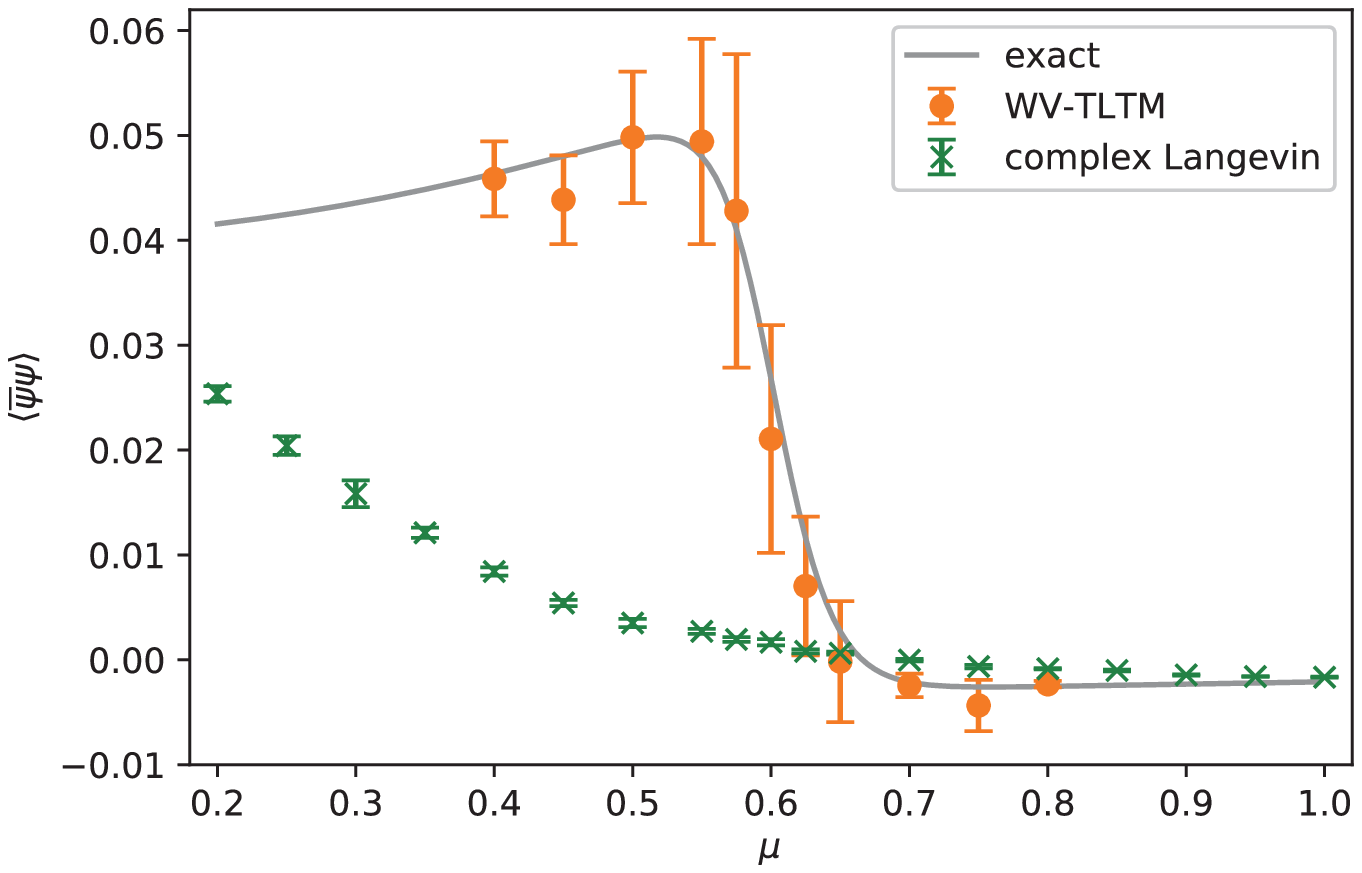}
  \hspace{5mm}
  \includegraphics[width=65mm]{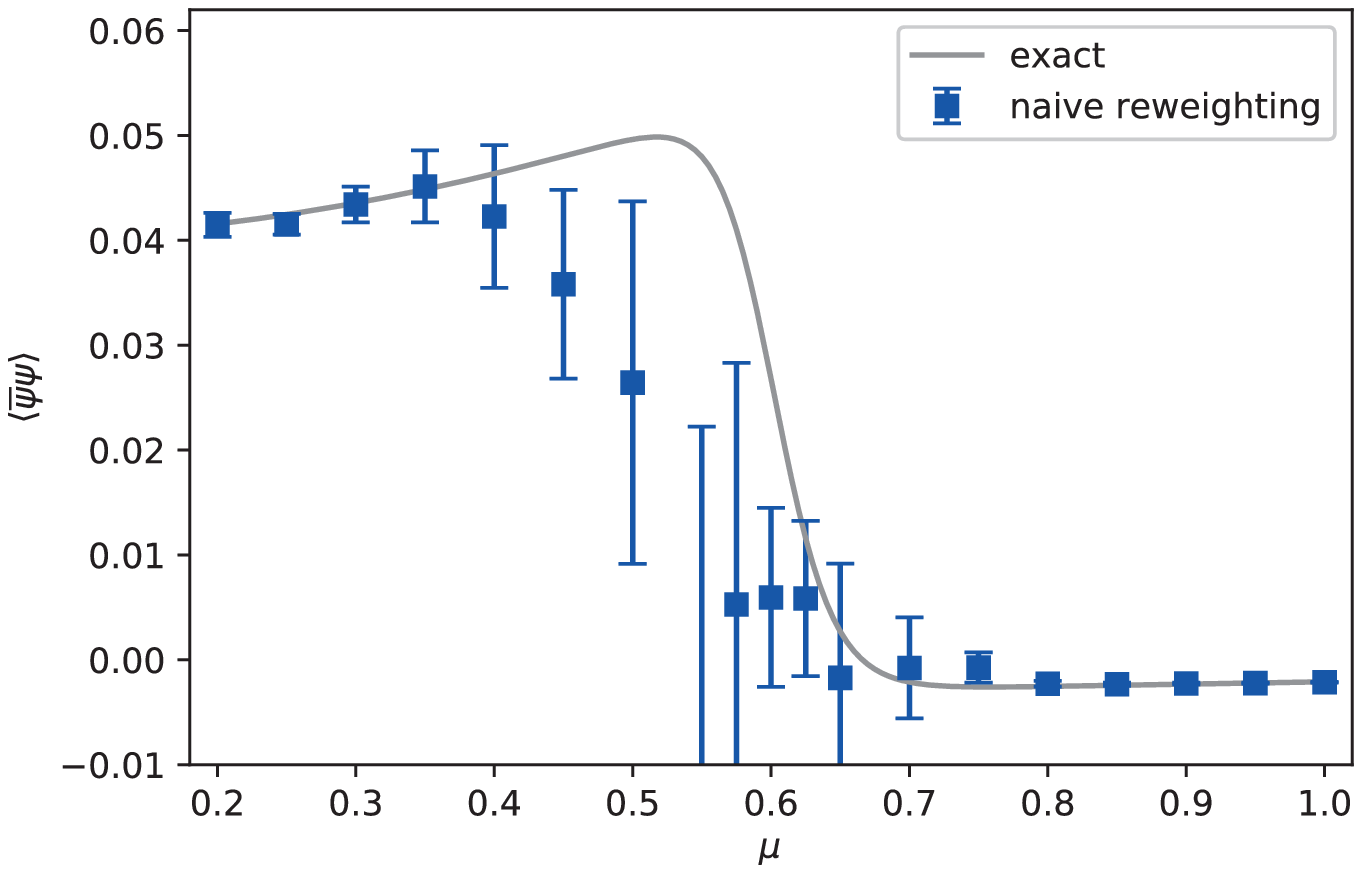} \\
  \includegraphics[width=65mm]{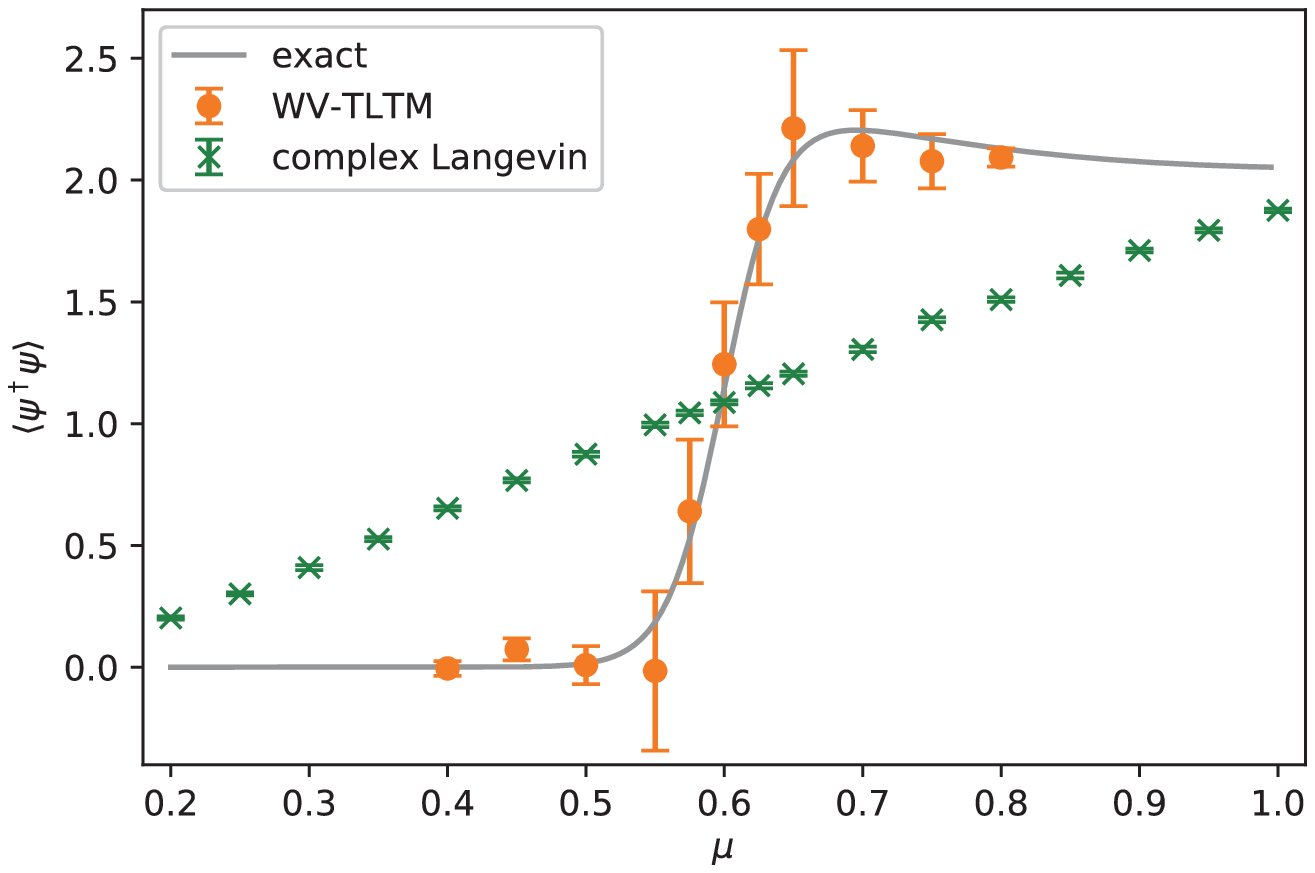}
  \hspace{5mm}
  \includegraphics[width=65mm]{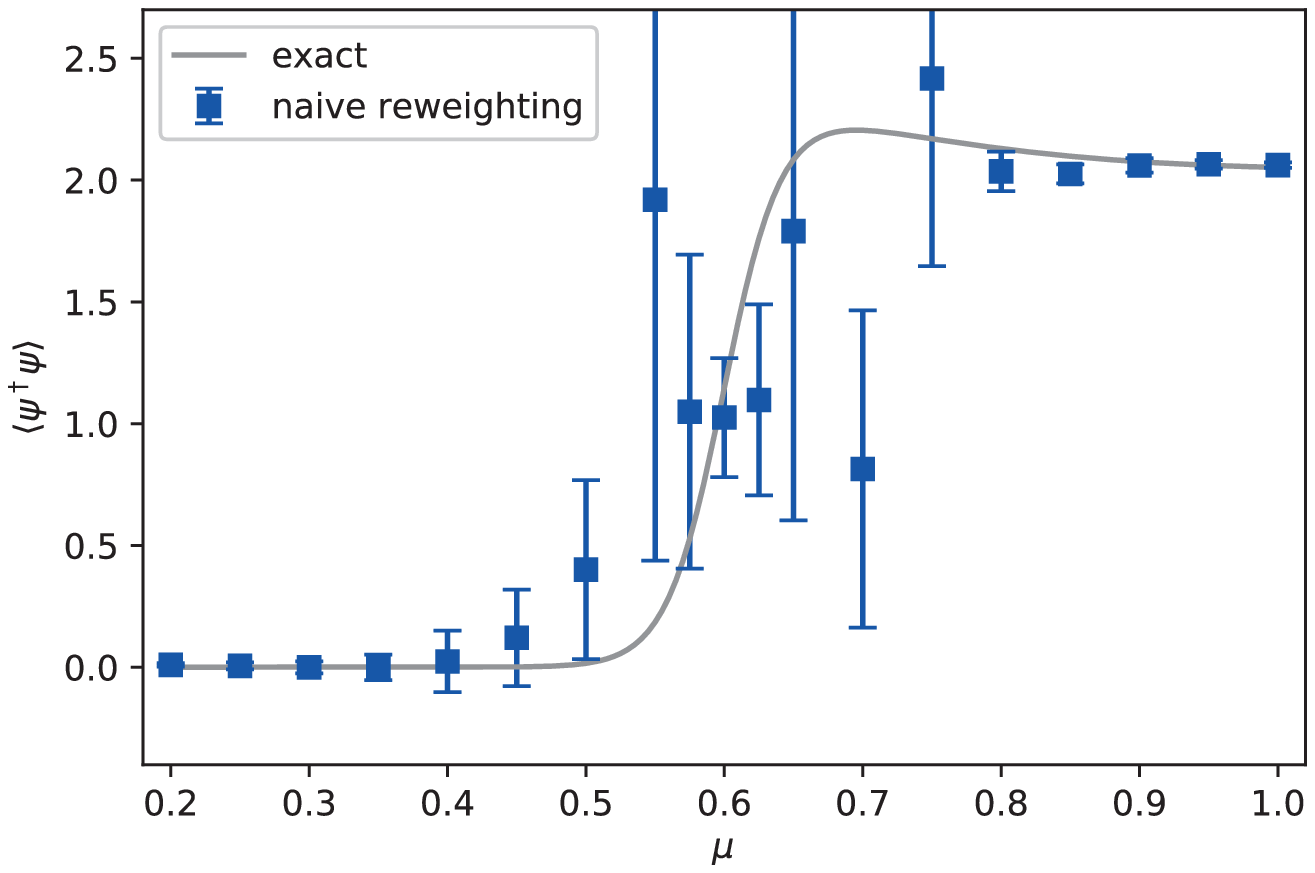}
  \caption{
    Estimates of the chiral condensate (upper panels) and the number density (lower panels) \cite{Fukuma:2020fez}.
  }
  \label{fig:wv_rw_cl}
\end{figure}

Figure~\ref{fig:wv_rw_cl} shows the results 
for varying $\mu$ with $n=10$ (thus $N=2n^2=200$) and $m=0.004$.
We also give for comparison 
the results obtained with the naive reweighting method 
and with the complex Langevin method 
(see \cite{Fukuma:2020fez} for detailed analysis). 
This size of matrix ($n=10$) 
requires more than about 70 replicas in the original TLTM, 
making the computation very costly. 
The sample size is set to 4,000--17,000 for the WV-TLTM varying on $\mu$, 
and to 10,000 for the naive reweighting and complex Langevin methods.

From the results with the naive reweighting, 
we see that there is a severe sign problem for this parameter.
We also see that the complex Langevin suffers from the wrong convergence 
(largely deviated means from the exact values with small statistical errors; 
see \cite{Bloch:2017sex} 
for a detailed analysis of the Stephanov model 
with the complex Langevin method). 
In contrast, the WV-TLTM gives results 
in good agreement with the exact values with controlled statistical errors, 
which proves the effectiveness of the WV-TLTM.

\section{Statistical analysis method for the WV-TLTM}
\label{sec:stat-analys-meth}

In generating configurations, 
$T_1$ must be set sufficiently large 
to reduce the sign problem and $T_0$ sufficiently small to recover ergodicity. 
In the estimation, however, 
we can use different values of $T_0,\,T_1$ 
(denoted by $\tilde{T}_0$,\, $\tilde{T}_1$) 
by restricting the data 
to those belonging to the interval $[\tilde T_0,\tilde T_1]$.
Note that in an ideal situation 
the estimate of a fixed observable 
does not depend on the choice of $\tilde{T}_0,\,\tilde{T}_1$ 
according to Cauchy's theorem. 
We can in turn use this fact in actual calculations 
to test the sufficiency of the sample size 
for the whole configuration space to be well explored. 
Furthermore, when necessary, 
we may exclude two extreme regions: 
one is the small $t$ region suffering from phase fluctuations, 
and the other is the large $t$ region 
having possible large autocorrelations and systematic errors 
due to the complicated geometry of $\Sigma_t$ at large $t$. 

We now show that 
the restricted data set can be regarded as obtained from a Markov chain 
\cite{Fukuma:2021aoo}. 
Let $P(z'|z)$ be 
the transition matrix in the whole worldvolume with $[T_0,T_1]$, 
and denote by $\tilde{\mathcal{R}}$ 
the restricted region with $[\tilde{T}_0$, $\tilde{T}_1]$. 
Then, the probability $\tilde{P}(z'|z)$ 
to obtain $z'\in \tilde{\mathcal{R}}$ 
from $z \in \tilde{\mathcal{R}}$ can be expressed as 
\begin{align}
  \tilde{P}(z'|z) = P(z'|z) +
  \int_{{\tilde \calr}^c} dw\, P(z'|w)P(w|z) + 
  \int_{{\tilde \calr}^c} dw_2 dw_1\, P(z'|w_2)P(w_2|w_1)P(w_1|z) + \cdots, 
  \label{eq:ptilde}
\end{align}
where $\tilde{\calr}^c\equiv\calr\backslash \tilde{\calr}$. 
The first term describes a direct transition in $\tilde{\calr}$, 
while the rest describes processes that leave $\tilde\calr$ once. 
The ergodicity of $P$ ensures that of $\tilde P$, 
so that 
$\tilde P$ has the correct normalization as a transition matrix. 
Furthermore, if $P$ satisfies the detailed balance 
with equilibrium distribution $\rho_{\rm eq}(z)$, 
$\tilde{P}$ also satisfies the detailed balance 
with the equilibrium distribution 
$\tilde\rho_{\rm eq}(z)\equiv
\rho_{\rm eq}(z)/\int_{\tilde\calr}dz\, \rho_{\rm eq}(z)$. 
We thus can apply standard statistical analysis methods 
(such as the Jackknife) to the restricted data set. 

\begin{figure}[t]
  \centering
  \includegraphics[width=65mm]{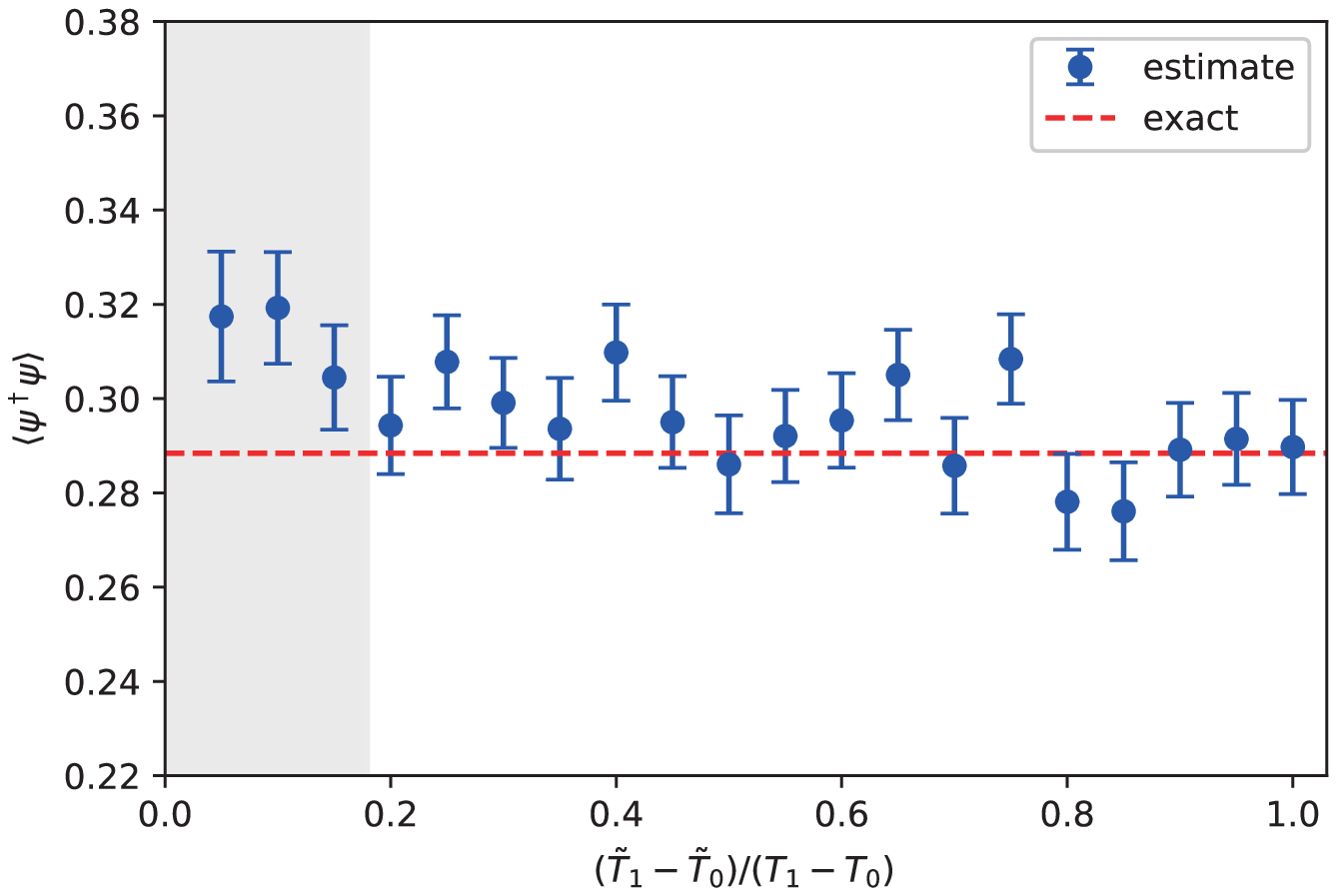}\\
  \includegraphics[width=65mm]{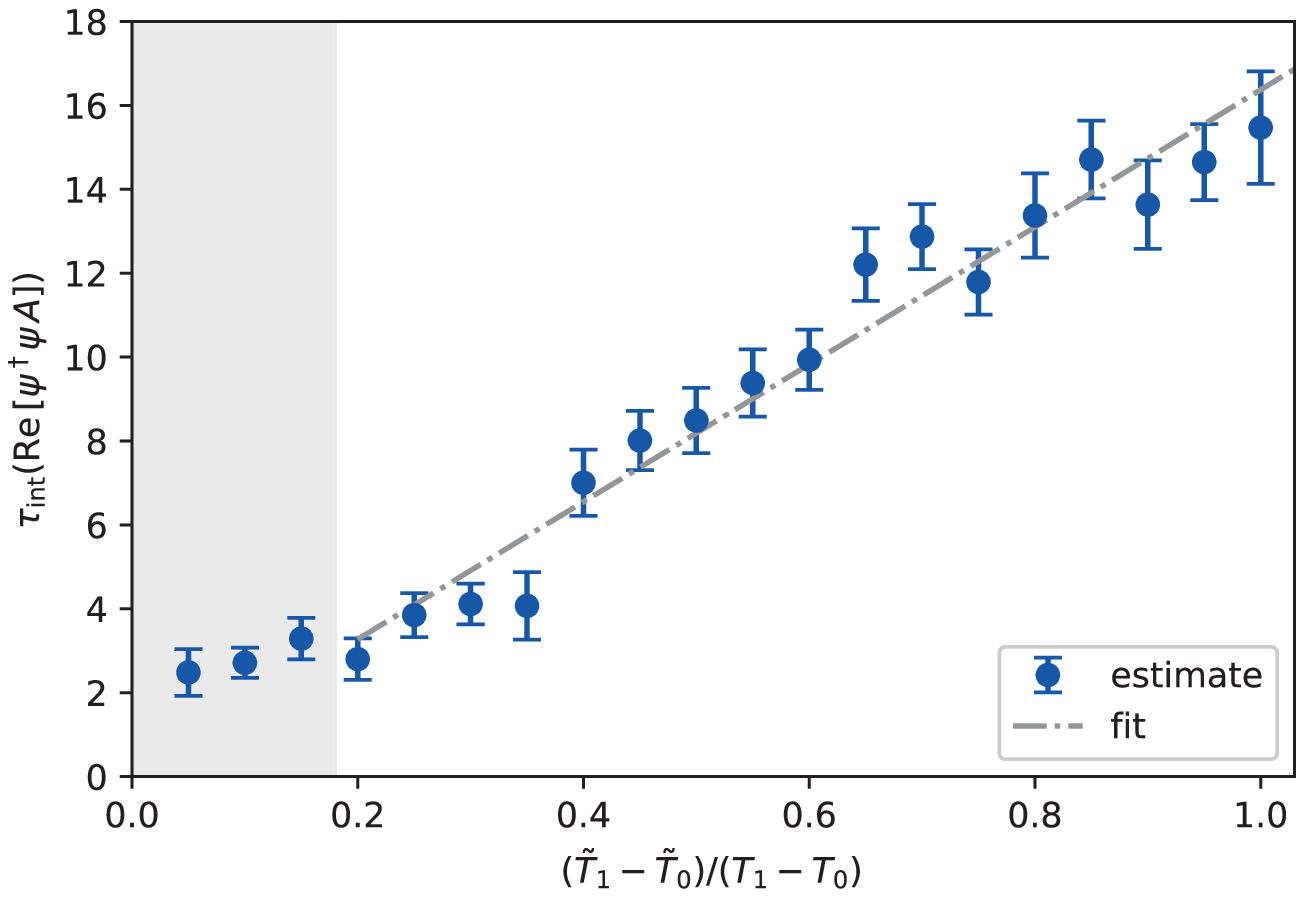}
  \hspace{6.8mm}
  \includegraphics[width=65mm]{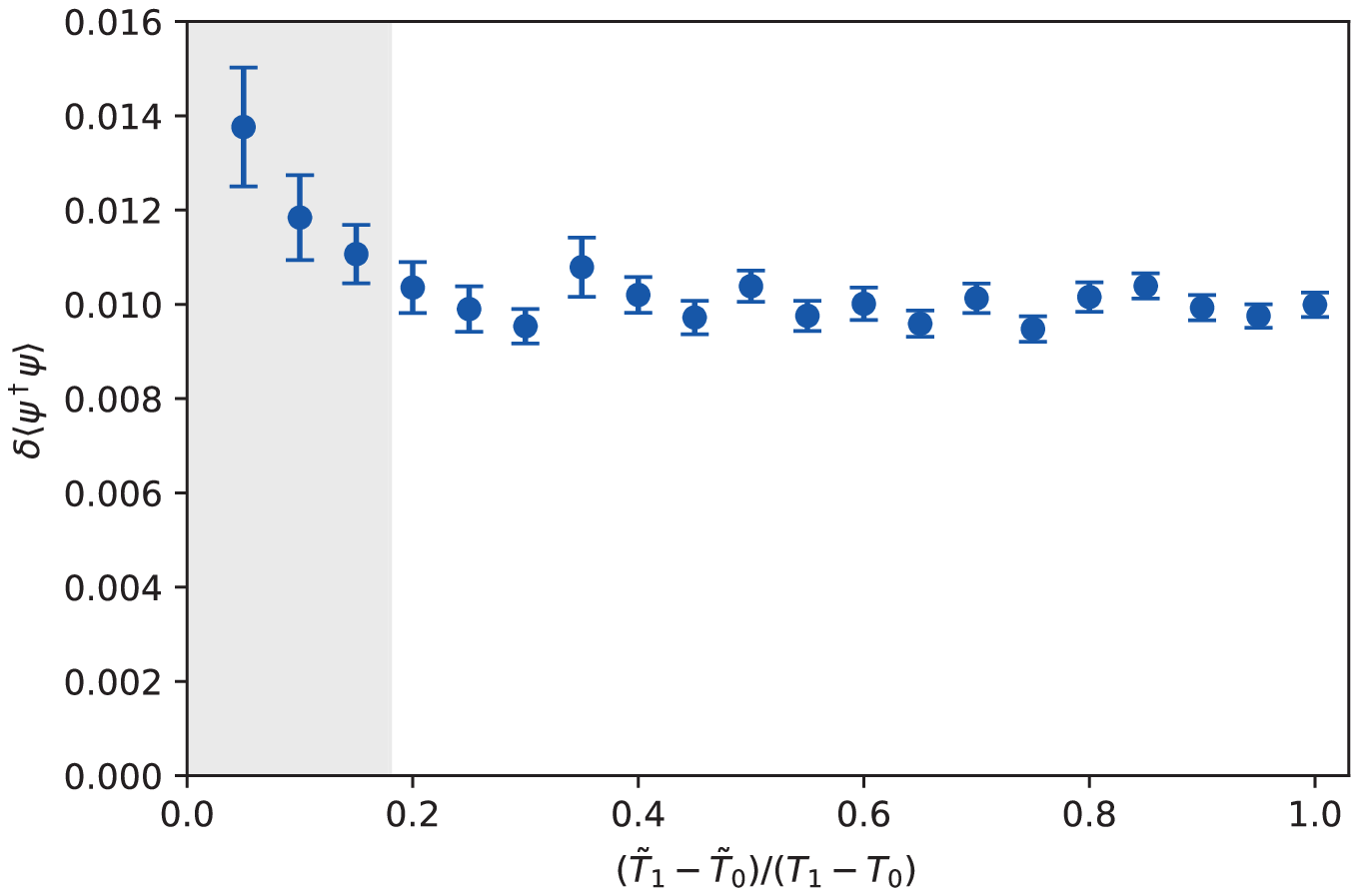}
  \caption{
    The dependence of statistical quantities 
    on the choice of $\tilde{\mathcal{R}}$ 
    with varying $\tilde T_0$ and fixed $\tilde T_1$. 
    (Top) $\tilde{T}_0$-independence of the estimates 
    \cite{Fukuma:2021aoo}. 
    (Bottom left) Scaling of $\tilde{\tau}_{{\rm int}}$, 
    Eq.~\eqref{eq:tauint_scaling}, 
    and (Bottom right) $\tilde{T}_0$-independence 
    of the statistical errors, 
    Eq.~\eqref{eq:delta_indep} 
    \cite{Fukuma:2021aoo}. 
    The integrated autocorrelation times are saturated in the shaded region, 
    $\tilde{\tau}_{{\rm int}} \simeq 1$, 
    where the scaling \eqref{eq:tauint_scaling} breaks down. 
  }
  \label{fig:expec_stat}
\end{figure}

Figure~\ref{fig:expec_stat} shows 
the estimates of the baryon number density 
with varying $\tilde{T}_0$ and fixed $\tilde{T}_1$ 
for the Stephanov model of a matrix size $n=2$ 
(see \cite{Fukuma:2021aoo} for details). 
We observe a plateau for a wide range of 
$r\equiv (\tilde{T}_1-\tilde{T}_0)/(T_1-T_0)$, 
from which we confirm that 
the estimates do not depend on $\tilde{T}_0$. 
The estimates are slightly deviated from the exact value 
at $r \lesssim 0.2$, 
but this can be understood by the smallness of the sample size 
when $\tilde{T}_0$ gets close to $\tilde{T}_1$. 

One may be worried about the increase in statistical errors 
due to the reduced size of restricted data set. 
We can argue that this is not the case 
because the effect of the decrease in the sample size 
is compensated by the decrease in autocorrelation. 
In fact, one can show that 
the effective sample size $N_{\rm conf}^{\rm eff}$ defined by
\begin{align}
  N_{\rm conf}^{\rm eff} \equiv N_{\rm conf}/\tau_{\rm int} 
\end{align}
does not depend on the choice of $\tilde{T}_0,\,\tilde{T}_1$. 
Here, $N_{\rm conf}$ is the original sample size, 
and $\tau_{\rm int}$ is the integrated autocorrelation time 
of a given operator $\calo$, 
defined from the autocorrelation functions 
$C_k\equiv 
\langle (\mathcal{O}_0-\langle \mathcal{O} \rangle )
( \mathcal{O}_k-\langle \mathcal{O} \rangle ) \rangle $
as 
\begin{align}
  \tau_{\rm int} \equiv 1 + 2\sum_{k=1}^{\infty} C_k / C_0. 
\end{align}
We assume that the flow time $t\in [T_0,T_1]$ is uniformly distributed, 
which is equivalent to assuming that 
the weight function $W(t)$ is ideally chosen 
[see discussion below Eq.~\eqref{eq:2}]. 
We also assume that the integrated autocorrelation time of $t$ is 1. 
Then, the probability for a configuration to appear 
in $[\tilde{T}_0,\tilde{T}_1]$ is given by 
\begin{align}
  p \equiv \frac{\tilde{T}_1-\tilde{T}_0}{T_1-T_0}. 
\end{align}
Since a configuration appears in the subregion $\tilde{\calr}$ 
with this probability, 
the integrated autocorrelation time of the subsample 
will be reduced as  
\begin{align}
  \tilde{\tau}_{\rm int} = p \tau_{\rm int}. 
\label{eq:tauint_scaling}
\end{align}
Since the sample size is reduced with the same ratio, 
$\tilde{N}_{\rm conf} = p N_{\rm conf}$, 
the effective sample size does not change:
\begin{align}
  \tilde{N}_{\rm conf}/\tilde{\tau}_{\rm int}
  = N_{\rm conf}/\tau_{\rm int} = N_{\rm conf}^{\rm eff}, 
\end{align}
which means that 
the statistical error $\delta\langle \calo \rangle$ 
to the estimate of $\langle \mathcal{O} \rangle$ 
is independent of the choice of $\tilde{T}_0,\tilde{T}_1$ 
because $\delta\langle \calo \rangle$ is given by the formula 
\begin{align}
  \delta\langle \calo \rangle = \sigma_{\calo} / \sqrt{N_{\rm conf}^{\rm eff}} 
  \quad 
  \bigl(\sigma_{\calo}^2\equiv\langle(\calo-\langle\calo\rangle)^2
  \rangle\bigr). 
\label{eq:delta_indep}
\end{align}
The scaling \eqref{eq:tauint_scaling} 
and the $\tilde{T}_0, \tilde{T}_1$-independence 
of Eq.~\eqref{eq:delta_indep} 
can actually be confirmed as in Fig.~\ref{fig:expec_stat}.

\section{Summary and outlook}
\label{sec:summary-outlook}

We have reported that 
the worldvolume tempered Lefschetz thimble method (WV-TLTM) works quite well, 
by demonstrating its effectiveness 
with the application to the Stephanov model. 
We also discussed the statistical analysis method for the WV-TLTM, 
which is expected to become useful 
especially when performing large-scale computations. 

We are now preparing large-scale simulations with the WV-TLTM 
for systems having the sign problem, 
such as finite density QCD, the Hubbard model, and real-time QM/QFTs. 
In parallel with research in this direction, 
we believe that 
it is still important to continue developing the algorithm itself, 
for example, to find an efficient algorithm 
to determine the weight function $W(t)$. 
A study along these lines is in progress and will be reported elsewhere. 

\acknowledgments
The authors thank Issaku Kanamori, Yoshio Kikukawa, and Jun Nishimura 
for useful discussions.
N.M. thanks Henry Lamm, Akio Tomiya, and Yukari Yamauchi 
for valuable discussions during the conference. 
This work was partially supported by JSPS KAKENHI 
Grant Numbers JP20H01900, JP18J22698, JP21K03553 
and by SPIRITS 
(Supporting Program for Interaction-based Initiative Team Studies) 
of Kyoto University (PI: M.F.). 
N.M.\ is supported by the Special Postdoctoral Researchers Program 
of RIKEN.



\end{document}